\documentclass[aps,prl,twocolumn,showpacs,10pt,superscriptaddress,preprintnumbers,nofootinbib]{revtex4-1}
\usepackage{epsfig,amssymb,amsmath,psfrag,epstopdf,color}
\pdfoutput=1
\usepackage{graphicx}
\usepackage{subfig}
\usepackage[margin=7pt,justification=centerlast]{caption}

\usepackage{paralist}
\usepackage{hyperref}

\allowdisplaybreaks

\def\beq{\begin{equation}}
\def\eeq{\end{equation}}
\def\bsp#1\esp{\begin{split}#1\end{split}}
\newcommand{\be}{\begin{equation}}
\newcommand{\ee}{\end{equation}}
\newcommand{\bea}{\begin{eqnarray}}
\newcommand{\eea}{\end{eqnarray}}

\def\Fig#1{Fig.~{\ref{#1}}}

\def\cO{{\mathcal O}}

\def\to{\rightarrow}

\def\ksl{\not{\hbox{\kern-2.3pt $k$}}}

\def\spa#1.#2{\left\langle#1\,#2\right\rangle}
\def\spb#1.#2{\left[#1\,#2\right]}
\def\lor#1.#2{\left(#1\,#2\right)}
\def\sand#1.#2.#3{%
\left\langle\smash{#1}{\vphantom1}^{-}\right|{#2}%
\left|\smash{#3}{\vphantom1}^{-}\right\rangle}

\def\cE{\mathcal{E}}

\DeclareRobustCommand{\Fig}[1]{Fig.~\ref{#1}}

\DeclareRobustCommand{\Eq}[1]{Eq.~(\ref{#1})}

\DeclareRobustCommand{\Refs}[1]{Refs.~\cite{#1}}

\newcommand{\nn}{\nonumber}

\begin{document}

\title{Conformal Colliders Meet the LHC}

\author{Kyle Lee}
\email{kylelee@lbl.gov}
\affiliation{Nuclear Science Division, Lawrence Berkeley National Laboratory, Berkeley, CA 94720}

\author{Bianka Me\c caj}
\email{bianka.mecaj@yale.edu}
\affiliation{Department of Physics, Yale University, New Haven, CT 06511}

\author{Ian Moult}
\email{ian.moult@yale.edu}
\affiliation{Department of Physics, Yale University, New Haven, CT 06511}

%%%%%%%%%%%%%%%%%%%%%%%%%%%
\begin{abstract}
The remarkably high energies of the Large Hadron Collider (LHC) have allowed for the first measurements of the shapes and scalings of multi-point correlators of energy flow operators, $\langle \Psi | \mathcal{E}(\vec n_1) \mathcal{E}(\vec n_2) \cdots \mathcal{E}(\vec n_k)  |\Psi \rangle$, providing new insights into the Lorentzian dynamics of quantum chromodynamics (QCD).
In this \emph{Letter}, we use recent advances in effective field theory to derive a rigorous factorization theorem for the light-ray density matrix, $\rho= |\Psi\rangle \langle \Psi |$, inside high transverse momentum jets at the LHC.
Using the light-ray operator product expansion, the scaling behavior of multi-point correlators can be computed from the expectation value of the twist-2 spin-$J$ light-ray operators, $\mathbb{O}^{[J]}$, in this state, $\text{Tr}[ \rho ~\mathbb{O}^{[J]} ]$.
We compute the light-ray density matrix at next-to-leading order, and combine this with results for the next-to-leading logarithmic scaling behavior of the correlators up to six-points, comparing with CMS Open Data. 
This theoretical accuracy allows us to resolve the quantum scaling dimensions of QCD light-ray operators inside jets at the LHC.
Our factorization theorem for the light-ray density matrix at the LHC completes the link between recent developments in the study of energy correlators and LHC phenomenology, opening the door to a wide variety of precision jet substructure studies. 
\end{abstract}
%%%%%%%%%%%%%%%%%%%%%%%%%%%

\maketitle

\emph{Introduction.}---The Large Hadron Collider (LHC) provides an opportunity to explore quantum field theory in general, and quantum chromodynamics (QCD) in particular, at unprecedented energy scales, and with modern resolution detectors \cite{CMSPF,ATLAS:2017ghe}. 
Due to the phenomenon of asymptotic freedom \cite{Gross:1973id,Politzer:1973fx,Gross:1973ju,Gross:1974cs}, this gain in energy is particularly advantageous, as it enables QCD to be studied in the perturbative regime, where first principles calculations are currently possible. 

One of the major achievements, which re-invigorated the study of QCD at the LHC, was the introduction of experimentally robust infrared safe jet algorithms, most notably the anti-$k_T$ algorithm \cite{Cacciari:2005hq,Salam:2007xv,Cacciari:2008gp,Cacciari:2011ma}, that allow for the identification of high transverse momentum, $p_T$, jets in hadronic collisions. 
The inclusive production of such jets can be studied using rigorous factorization theorems \cite{Collins:1981ta,Bodwin:1984hc,Collins:1985ue,Collins:1988ig,Collins:1989gx,Collins:2011zzd,Nayak:2005rt}, whose perturbative components have been computed to next-to-next-to-leading accuracy \cite{Gehrmann-DeRidder:2013uxn,Currie:2013dwa,Currie:2016bfm,Currie:2017eqf,Currie:2017ctp,Currie:2018xkj,Czakon:2019tmo}, allowing for precision studies of QCD in hadron colliders. 

However, many of the most fascinating questions about QCD, namely understanding the Lorentzian dynamics of quarks and gluons, and the nature of their real-time confinement into hadrons, are not encoded in the distribution of jets, but rather in the structure of energy flow {\emph{within}} jets, known as jet substructure \cite{Larkoski:2017jix,Marzani:2019hun}. 
Jet substructure has been extraordinarily successful as a new way to search for physics beyond the Standard Model \cite{Butterworth:2008iy,Kaplan:2008ie,Krohn:2009th}, and provides new opportunities to study the dynamics of QCD both in vacuum and medium \cite{Andrews:2018jcm,Cunqueiro:2021wls}. 

%%%%%%%%%%%%%%%%%%%%%%%%%%%%%%%
\begin{figure}
\includegraphics[width=0.955\linewidth]{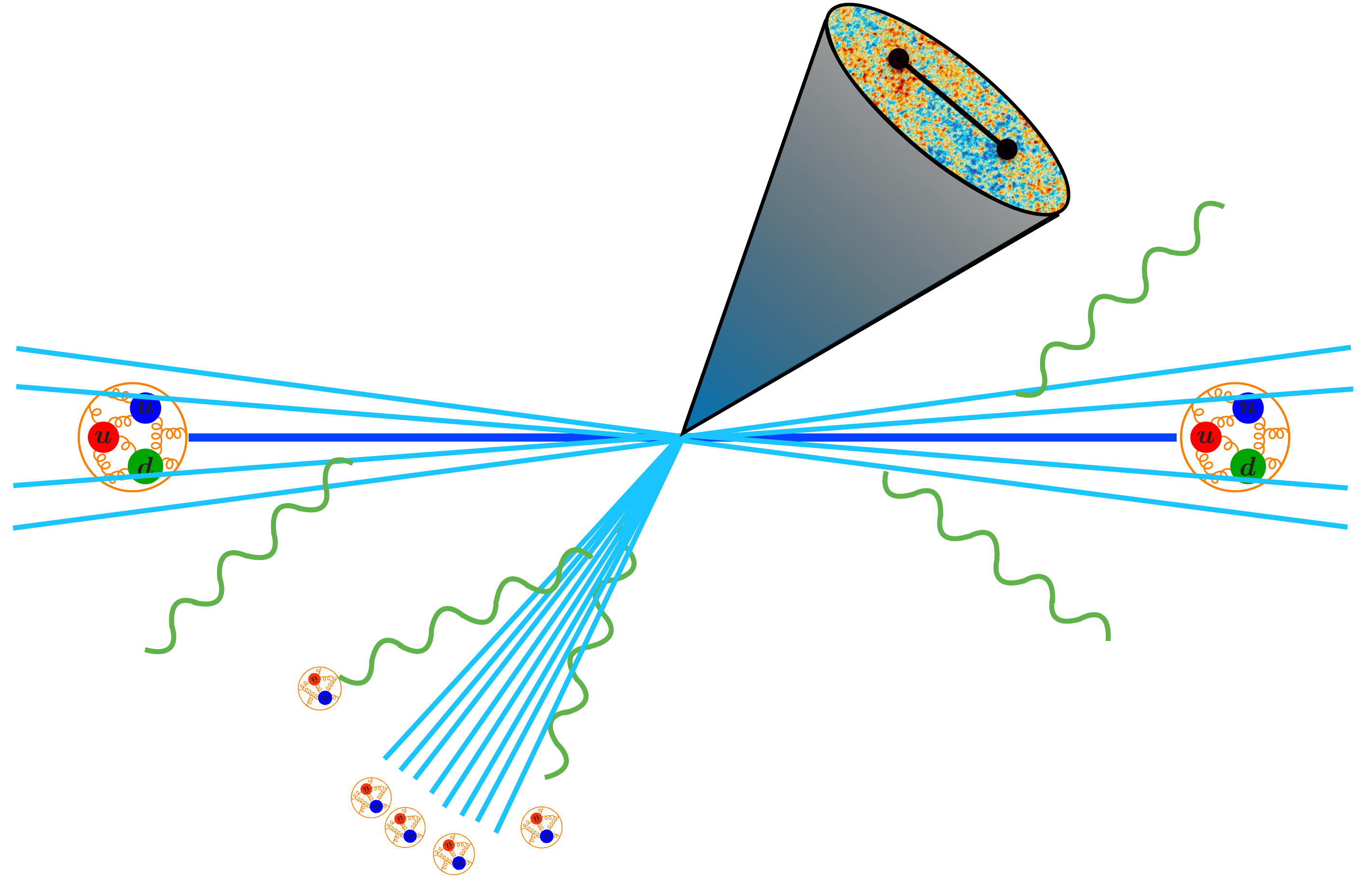}
\caption{An illustration of the two-point correlation function, $\langle \Psi | \cE(\vec n_1) \cE(\vec n_2)  |\Psi \rangle$, measured inside high-$p_T$ jets produced in collisions at the LHC.
}
\label{fig:jet_schematic}
\end{figure}
%%%%%%%%%%%%%%%%%%%%%%%%%%%%%%%

From a theoretical perspective, jet substructure is the study of the statistical properties of the asymptotic energy flux in collider experiments. 
It was placed in a modern field theoretic context in \cite{Hofman:2008ar}, where it was shown that it can be formulated as the study of correlation functions of $\langle \Psi | \cE(\vec n_1) \cE(\vec n_2) \cdots \cE(\vec n_k)  |\Psi \rangle$ of particular light-ray (ANEC) operators
\cite{Sveshnikov:1995vi,Tkachov:1995kk,Korchemsky:1999kt,Bauer:2008dt,Hofman:2008ar,Belitsky:2013xxa,Belitsky:2013bja,Kravchuk:2018htv}
\begin{align}\label{eq:ANEC_op}
\mathcal{E}(\vec n) = \lim_{r\to \infty}  \int\limits_0^\infty dt~ r^2 n^i T_{0i}(t,r \vec n)\,,
\end{align}
measured inside high energy jets at the LHC, as illustrated in \Fig{fig:jet_schematic}. For simplicity, we will occasionally use a shorthand notation $\cE(\vec n_i)\to \cE_i$.

An improved understanding of the properties of ANEC operators has been central to many recent advances in diverse areas of QFT, ranging from constraining conformal field theory (CFT) data \cite{Hofman:2008ar,Zhiboedov:2013opa,Cordova:2017zej,Cordova:2017dhq,Manenti:2019kbl}, to energy inequalities \cite{Bousso:2015wca,Hartman:2016lgu,Balakrishnan:2017bjg,Ceyhan:2018zfg}, to asymptotic symmetries \cite{Cordova:2018ygx,Korchemsky:2021htm} and the study of entropy in QFT \cite{Faulkner:2016mzt,Bousso:2016vlt,Casini:2017roe,Leichenauer:2018obf,Balakrishnan:2019gxl}. 
Excitingly, reformulating jet substructure such that it can draw from these diverse areas has led to significant recent progress. 
See e.g. \cite{Hofman:2008ar,Belitsky:2013xxa,Belitsky:2013bja,Belitsky:2013ofa,Korchemsky:2015ssa,Belitsky:2014zha,Dixon:2018qgp,Luo:2019nig,Henn:2019gkr,Chen:2019bpb,Dixon:2019uzg,Korchemsky:2019nzm,Chicherin:2020azt,Kravchuk:2018htv,Kologlu:2019bco,Kologlu:2019mfz,Chang:2020qpj,Dixon:2019uzg,Chen:2020uvt,Chen:2020vvp,Chen:2019bpb,Chen:2020adz,Chicherin:2020azt,Chen:2021gdk,Korchemsky:2021okt,Korchemsky:2021htm,Chang:2022ryc,Chen:2022jhb,Yan:2022cye}.
In particular, this has enabled multi-point correlation functions of ANEC operators to be studied using the light-ray operator product expansion (OPE) \cite{Hofman:2008ar,Kologlu:2019mfz,Chang:2020qpj} and celestial blocks \cite{Kologlu:2019mfz,Chang:2020qpj,Chen:2022jhb,Chang:2022ryc}, much in analogy with the study of multi-point correlation functions of local operators.

In the last several years there has been a program \cite{Chen:2020vvp} to bridge the gap between these formal developments and real world phenomenology.\footnote{See in particular \cite{Dixon:2019uzg,Chen:2019bpb,Chen:2020vvp,Chen:2020adz,Chen:2021gdk,Komiske:2022enw,Holguin:2022epo,Chen:2022jhb,EECgaussian}.} 
The goal of this program is to enable new theoretical techniques to be applied to real world collider data, extending our understanding of jet substructure, and conversely, to enable field theoretically interesting multi-point correlators of light-ray operators to be studied experimentally for the first time \cite{Komiske:2022enw,EECgaussian}.
In this \emph{Letter}, we take the final step, using recent developments in effective field theory (EFT) to derive rigorous factorization theorems allowing calculations of energy correlators to be seamlessly embedded into the complex LHC environment. 

%%%%%%%%%%%%%%%%%%%%%%%%%%%%%%%%%%%%%%%%%%%%%%%%
%%%%%%%%%%%%%%%%%%%%%%%%%%%%%%%%%%%%%%%%%%%%%%%%
\emph{The Light-Ray Operator Product Expansion and Light-Ray Densities}---The leading scaling behavior of multi-point correlators of energy flow operators in the small angle limit, as is relevant for the study of energy flow within jets at the LHC, is captured by the leading twist operators appearing in the light-ray OPE. 
At weak coupling, and for unpolarized jets, these are the twist-2 spin-$J$ light-ray operators  \cite{Hofman:2008ar,Chen:2020adz}
\begin{align}
\vec{\mathbb{O}}^{[J]}=\left( \mathbb{O}_q^{[J]}, \mathbb{O}_g^{[J]}  \right)^T=\lim_{r\to \infty} r^2 \int\limits_0^{\infty} dt\, \vec{\cO}^{[J]}(t,r\vec{n})\,,
\end{align}
obtained from the light transform \cite{Kravchuk:2018htv} of the standard twist-2 quark and gluon operators
\begin{align}
\cO_q^{[J]}&=\frac{1}{2^J}\bar \psi \gamma^+ ( iD^+)^{J-1}\psi\,, \nn \\
\cO_g^{[J]}&=-\frac{1}{2^J} F_a^{\mu +}(i D^+)^{J-2}F_a^{\mu +}\,.
\end{align}
Performing the iterated OPE of light-ray operators, one finds that at leading twist \cite{Hofman:2008ar,Chen:2021gdk}\footnote{Strictly speaking, beyond leading logarithmic order in a non-conformal theory, lightray operators with spins that differ from $J=k+1$ by terms proportional to the $\beta$-function appear in the OPE \cite{Dixon:2019uzg}. These are taken into account in our analysis, but are suppressed here for simplicity. }
\begin{align} \label{eq:general_OPE}
\mathcal{E}(\vec{n}_1)\mathcal{E}(\vec{n}_2)\cdots  \mathcal{E}(\vec{n}_k)
&= \frac{1}{R_L^2} \left\{ f^{[k]}_q(u_i,v_i) \mathbb{O}_q^{[k+1]}(\vec{n}_1)   \right.  \\
&\left.+ f^{[k]}_g(u_i,v_i) \mathbb{O}_g^{[k+1]}(\vec{n}_1) \right \} +\mathcal{O}(R_L^0) \,, \nn
\end{align}
where $R_L$ is a scaling variable\footnote{For applications at hadron colliders, we use boost-invariant coordinates $\Delta R^2= \Delta \phi^2 +\Delta \eta^2$, with $\phi$ the azimuthal angle, and $\eta$ the rapidity.}, which for concreteness we take to be the largest angle between the energy flow operators, and $f_j^{[k]}(u_i, v_i)$ are non-trivial functions of the cross ratios for $k>2$. 

The light-ray OPE reduces the study of multi-point correlation functions to the study of the matrix elements $ \langle \mathbb{O}_i^{[J]}(\vec{n}_1) \rangle= \langle \Psi | \mathbb{O}_i^{[J]}(\vec{n}_1) | \Psi \rangle $, where $| \Psi \rangle$ is the state for an identified jet at the LHC. 
Formally, this state can be described as being sourced by a soft collinear effective theory (SCET) quark or gluon field \cite{Bauer:2000ew, Bauer:2000yr, Bauer:2001ct, Bauer:2001yt, Bauer:2002nz}.

%%%%%%%%%%%%%%%%%%%%%%%%%%%%%%%
\begin{figure}
\includegraphics[width=0.955\linewidth]{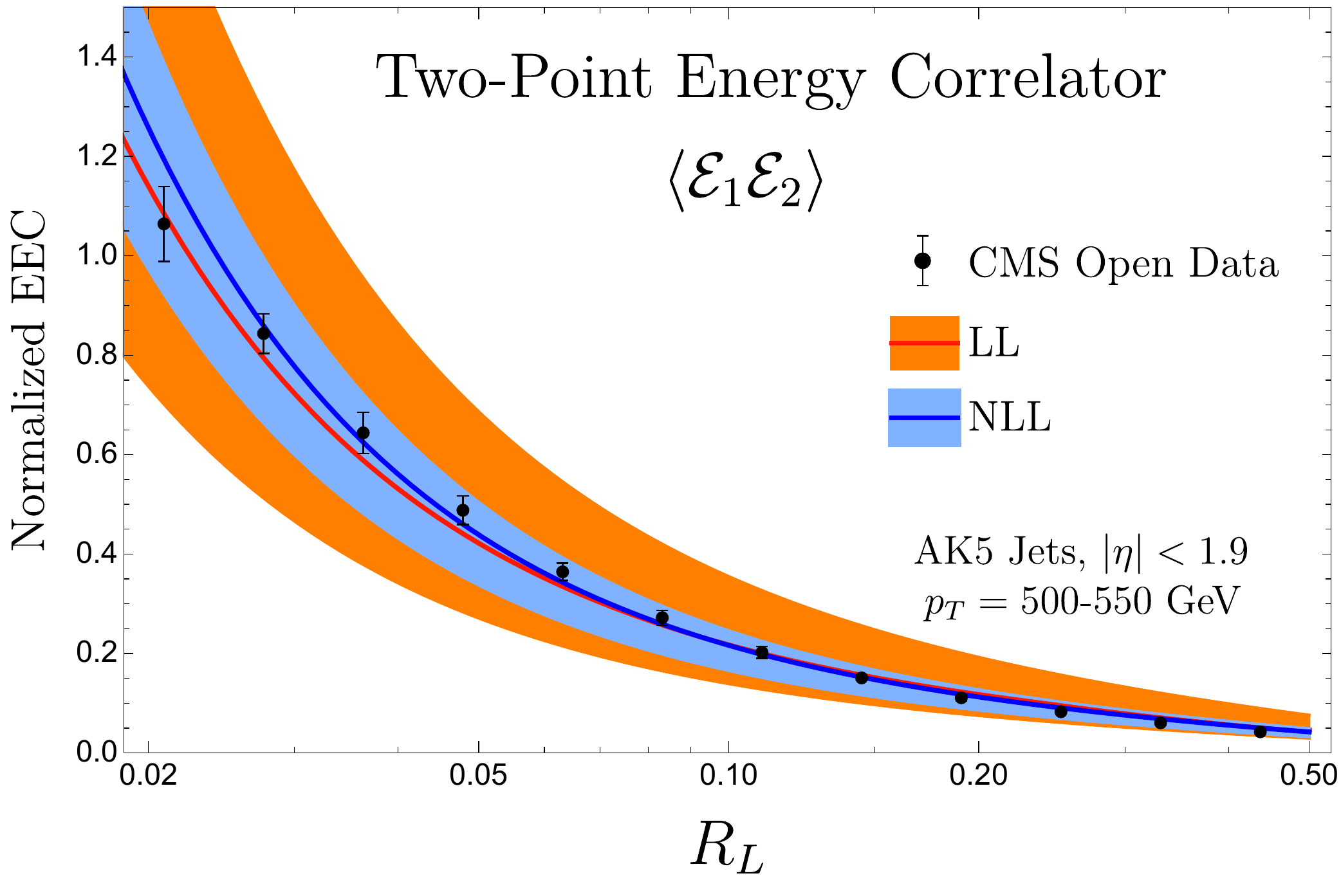}
\caption{A comparison of the LL and NLL predictions for the normalized two-point correlator, $\langle \cE(\vec n_1) \cE(\vec n_2) \rangle$, with CMS Open Data points extracted from \cite{Komiske:2022enw}.}
\label{fig:EEC}
\end{figure}
%%%%%%%%%%%%%%%%%%%%%%%%%%%%%%%

It is useful to rewrite this expectation value in terms of the so called  light-ray density matrix $\rho=| \Psi \rangle \langle \Psi |$ \cite{Chang:2022ryc},  as $ \langle \mathbb{O}_i^{[J]} \rangle =\text{Tr} [ \rho\, \mathbb{O}_i^{[J]}]$. 
Much like for the light-ray operators, it is convenient to work with states of definite quantum numbers under the Lorentz group, namely the celestial quantum numbers $\delta$ associated with boosts, and $j$ associated with transverse rotations $| \Psi \rangle=\sum_{\delta, j} |\Psi_{\delta, j} \rangle$. 
A primary complexity of jet physics is that standard observables are sensitive to the infinite sum over states of arbitrary quantum numbers. 
This has led to different approaches, such as grooming \cite{Dasgupta:2013ihk,Larkoski:2014wba} to attempt to simplify it.  
The energy correlators allow for a far more dramatic simplification by exploiting the underlying symmetries of the theory. 
The measurement of $\mathbb{O}_i^{[J]}$ projects onto the single state with definite quantum numbers  $|\Psi_{\delta=J+1, j=0} \rangle$. 
Measurements of the low point correlators project onto the particularly simple low $\delta$ states, ultimately leading to their simple theoretical structure. 
The OPE has therefore reduced the calculation of multi-point correlators at the LHC, to computing the probability to produce jet states $|\Psi_{\delta, j} \rangle$. 
This represents a clear departure from the way in which jets are more traditionally studied at the LHC, heavily motivated by developments in CFT.

%%%%%%%%%%%%%%%%%%%%%%%%%%%%%%%%%%%%%%%%%%%%%%%%%%%%%%%%%%%%%%
\emph{Factorization Theorem for Light-Ray Densities.}---A factorization theorem for the energy correlators in the collinear limit in $e^+e^-$ colliders was derived in \cite{Dixon:2019uzg}, allowing them to be systematically resummed at higher perturbative orders.\footnote{At leading logarithmic order, one can use the jet calculus \cite{Konishi:1979cb}. However, this does not allow one to take into account higher order corrections associated with the source.} 
The main complexity in extending this factorization theorem to the case of hadron colliders is the need to measure the energy correlators inside an identified jet defined with the anti-$k_T$ algorithm \cite{Cacciari:2008gp} used experimentally. 
The derivation of such a factorization theorem is the main result of this \emph{Letter}.

%%%%%%%%%%%%%%%%%%%%%%%%%%%%%%%
\begin{figure}
\includegraphics[width=0.955\linewidth]{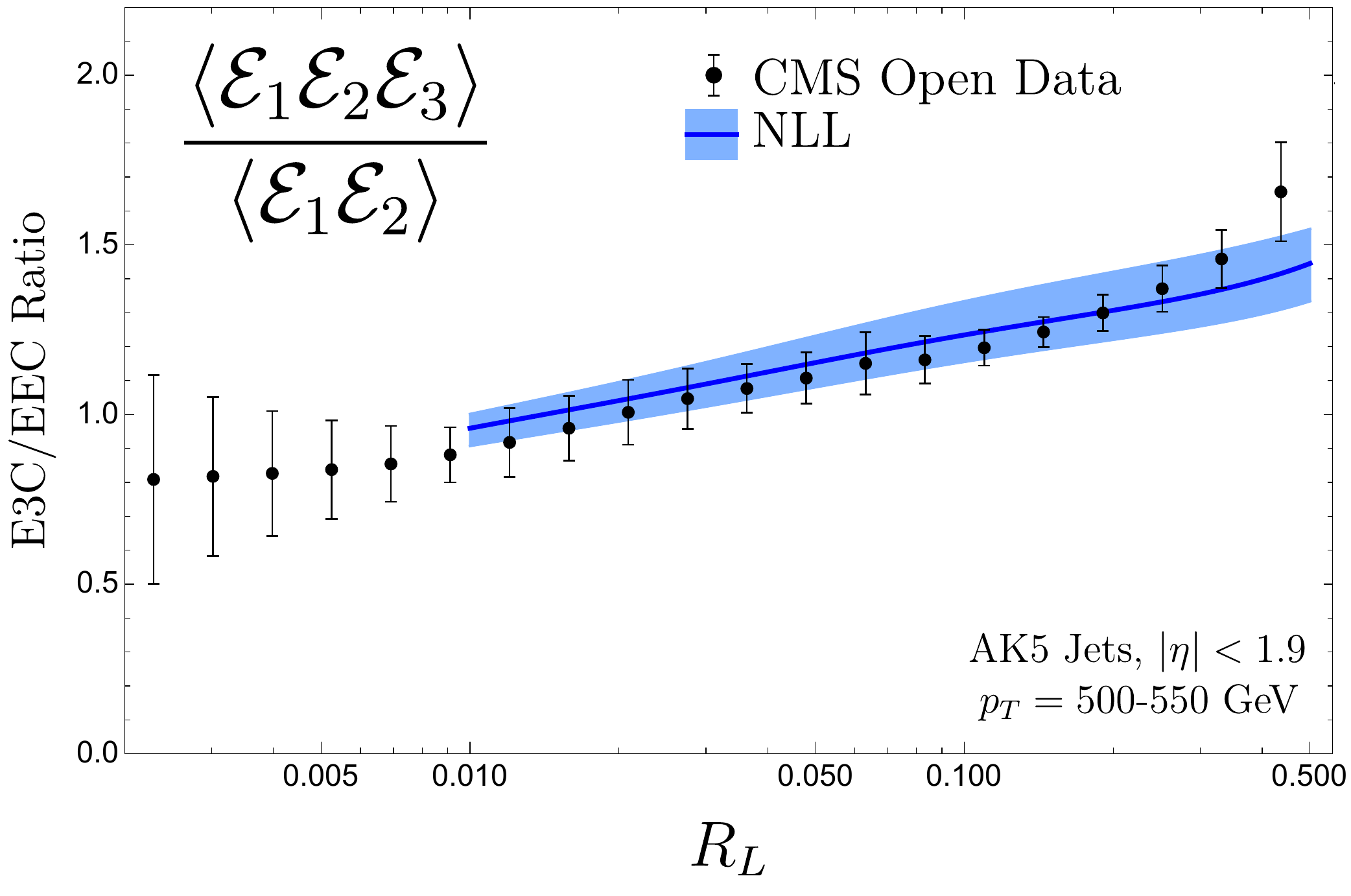}
\includegraphics[width=0.955\linewidth]{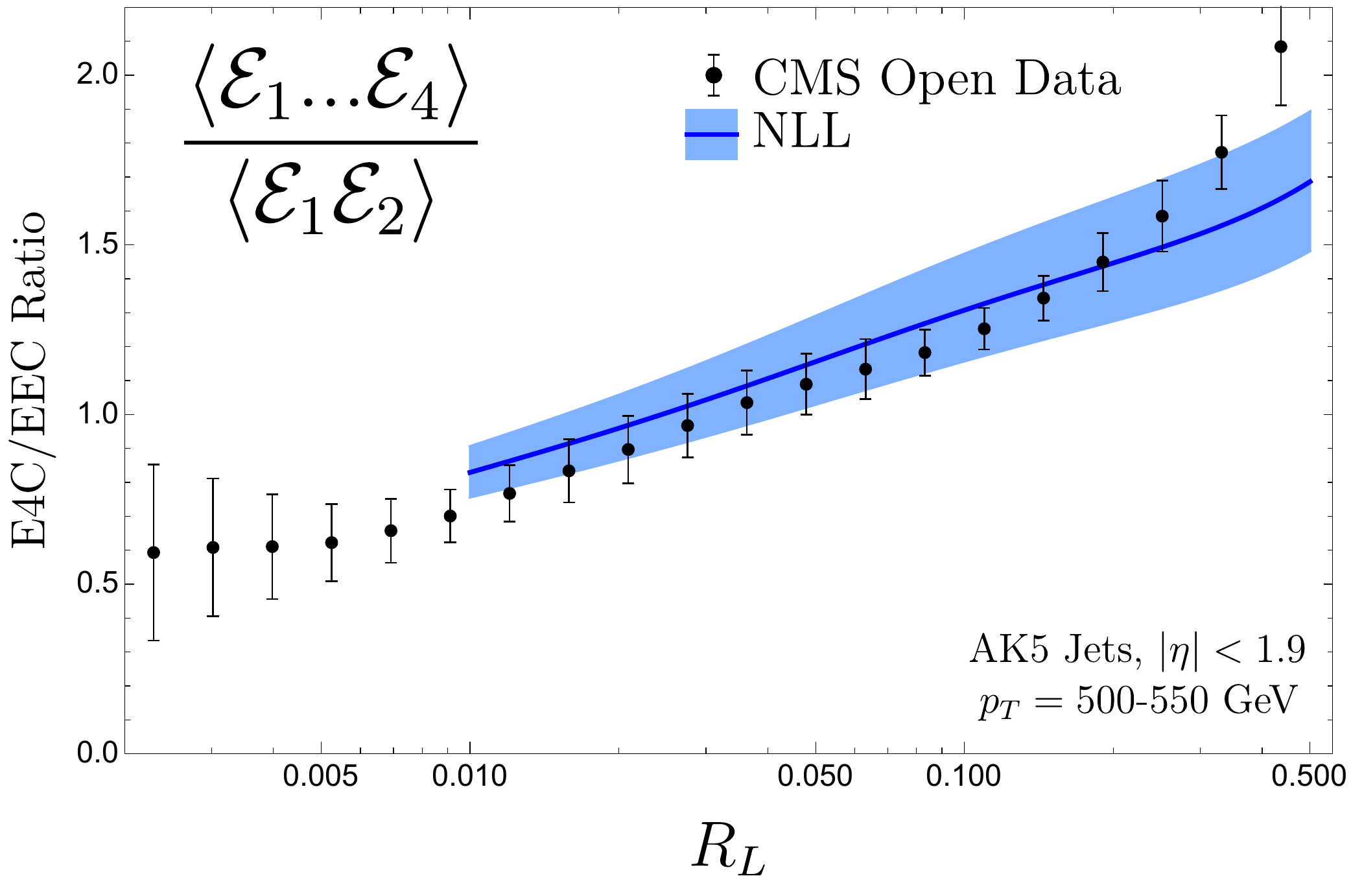}
\includegraphics[width=0.955\linewidth]{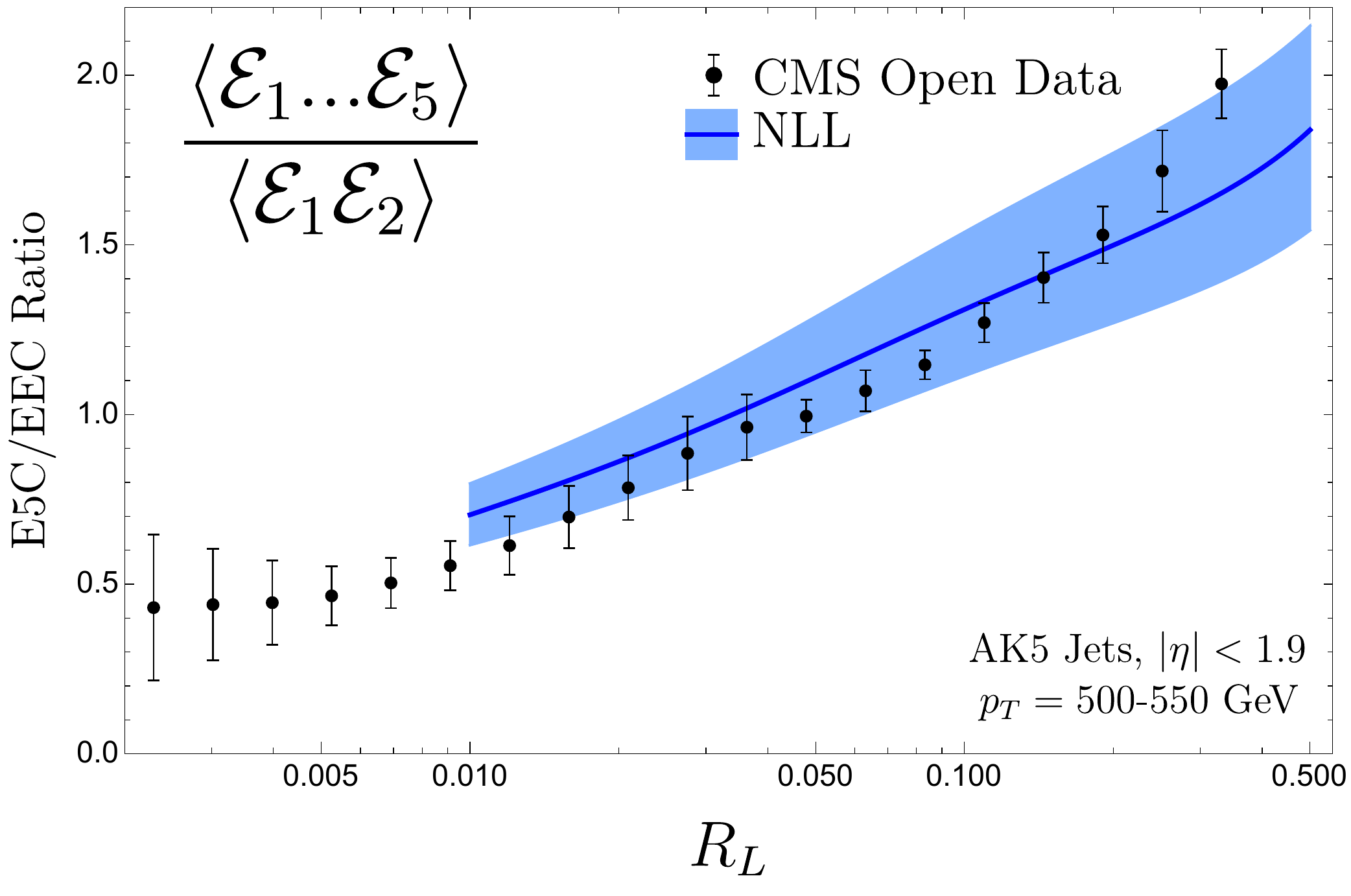}
\includegraphics[width=0.955\linewidth]{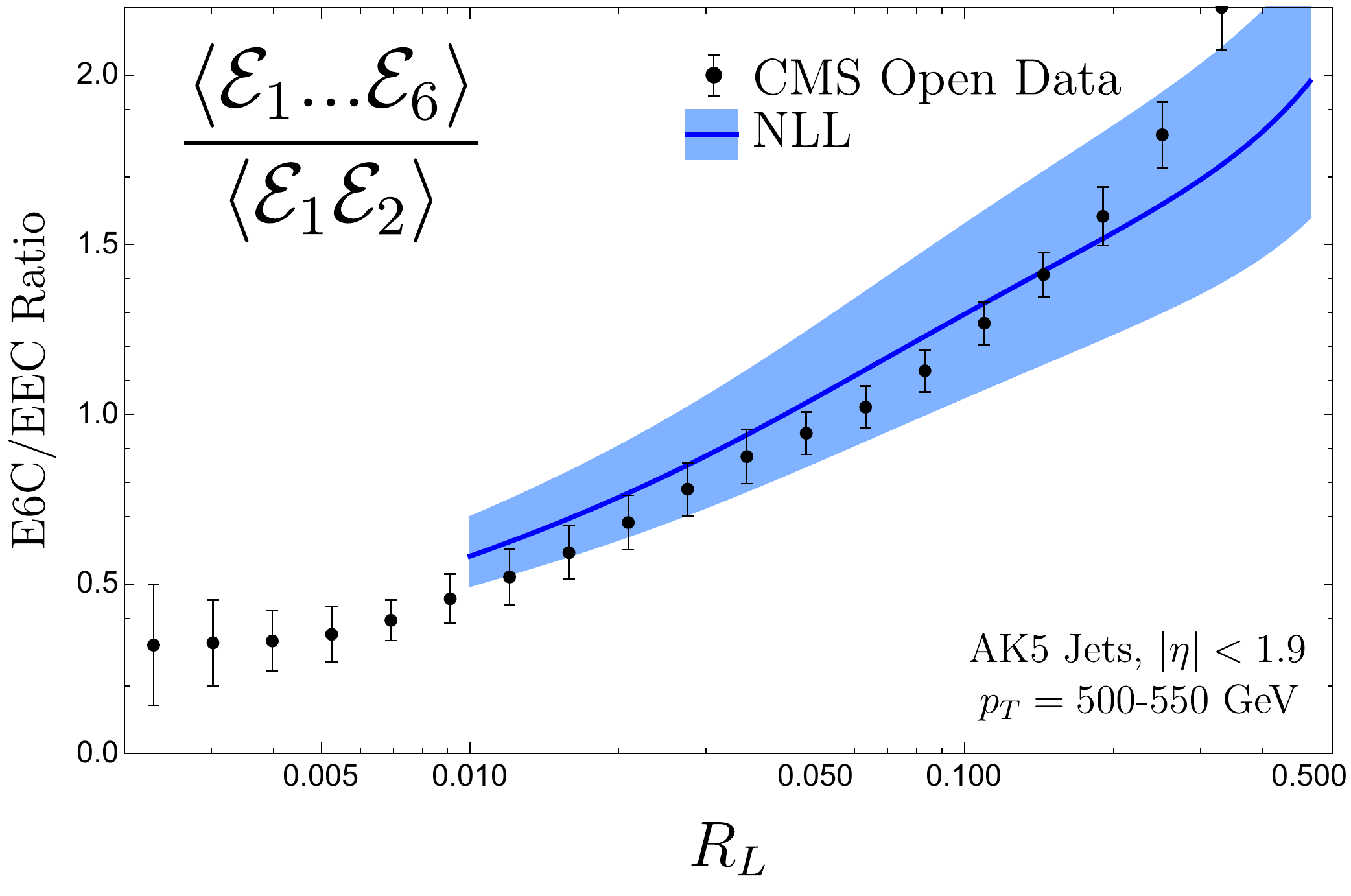}
\caption{Ratios of the multi-point projected correlators compared with CMS Open Data points extracted from \cite{Komiske:2022enw}. These ratios probe the quantum scaling associated with higher spin light-ray operators, $\mathbb{O}^{[J]}$, providing a quantitative test of the light-ray OPE.
}
\label{fig:EEC_ratios}
\end{figure}
%%%%%%%%%%%%%%%%%%%%%%%%%%%%%%%

Our factorization theorem is derived by combining the seminal work of Collins-Soper-Sterman \cite{Collins:1981ta,Bodwin:1984hc,Collins:1985ue,Collins:1988ig,Collins:1989gx,Nayak:2005rt,Collins:2011zzd} on factorization for inclusive hadron production, with recent developments in SCET \cite{Bauer:2000ew, Bauer:2000yr, Bauer:2001ct, Bauer:2001yt, Bauer:2002nz}, in particular the fragmenting jets formalism \cite{Procura:2009vm,Jain:2011xz,Procura:2011aq,Kang:2016mcy,Kang:2016ehg,Kang:2020xyq}, which allows for the factorization of states produced inside high $p_T$ jets.\footnote{For other phenomenological applications of the fragmenting jet function formalism see e.g. \cite{Liu:2017pbb,Kang:2017frl,Kang:2018jwa,Liu:2018ktv,Kang:2018qra,Kang:2018vgn,Aschenauer:2019uex,Cal:2021fla}.}
Combining these approaches, we derive a factorization theorem for generic correlation functions of energy flow operators inside inclusive high-$p_T$ jets at the LHC
\begin{align}\label{eq:fact_theorem}
\frac{\mathrm{d} \Sigma}{\mathrm{d} p_{T}\,\mathrm{d} \eta\, \mathrm{d} \{\zeta\}}
=& \sum_{i} \mathcal{H}_{i}\left(p_{T}/z,\eta, \mu\right)\\
&\hspace{-0.5cm}\otimes \int_0^1 dx\, x^N\,\mathcal{J}_{ij}(z,x,p_T R,\mu)\,J^{[N]}_j(\{\zeta\}, x,\mu)\,. \nn
\end{align}
Here, $\mathcal{H}_i$ is the hard function describing the production of parton $i$, which includes the parton distribution functions of the hadron beams. 
The convolution structure $\otimes$ is over the parton momentum fraction $z$. 
The matching coefficient $\mathcal{J}_{ij}$ incorporates the details of the jet algorithm~\cite{Kang:2016ehg}, and depends explicitly on the jet radius, $R$. 
Combined,  $\mathcal{H}_i$ and $\mathcal{J}_{ij}$ the light-ray density.
The energy correlator jet function~\cite{Dixon:2019uzg,Chen:2020vvp}, $J^{[N]}_j$, depends on a momentum fraction $x$, and the set of angles between the detectors, $\{\zeta\}$, which can be written in terms of cross ratios and $R_L$, as in \Eq{eq:general_OPE}. 
This factorization theorem enables any calculation of the energy correlators in the collinear limit to be embedded into the LHC environment. 
A particularly beautiful aspect of this factorization is that each component satisfies a DGLAP \cite{Gribov:1972ri,Dokshitzer:1977sg,Altarelli:1977zs} evolution, illustrating a natural interplay of the energy correlator and fragmenting jet formalisms.

As compared with previous approaches to jet substructure, we believe that the approach of correlation functions, encapsulated in \Eq{eq:fact_theorem}, presents a number of advantages. 
First, for an arbitrary $N$-point correlator, both the normalization and the shape dependence are free of soft effects and non-global logarithms \cite{Dasgupta:2001sh}, \emph{without} the application of grooming procedures \cite{Dasgupta:2013ihk,Larkoski:2014wba} that complicate higher order calculations. 
Second, unlike for groomed observables, where the algorithm complicates the non-perturbative structure of the observable \cite{Frye:2016okc,Frye:2016aiz,Hoang:2017kmk,Hoang:2019ceu}, all algorithm dependence lies in the matching coefficient, $\mathcal{J}_{ij}$, simplifying the structure of non-perturbative corrections \cite{Korchemsky:1995zm,Korchemsky:1994is,Korchemsky:1997sy,Korchemsky:1999kt,Belitsky:2001ij}. 
Third, since the inclusive hadron hard function is known at next-to-next-to-leading order (NNLO) \cite{Czakon:2021ohs}, and the fragmenting jet functions can be computed at NNLO using the approach of \cite{Liu:2021xzi}, the path to higher perturbative orders is clear. 
Finally, correlation functions of energy flow operators can be computed on tracks using the track function formalism for energy correlators \cite{Chang:2013rca,Chang:2013iba,Jaarsma:2022kdd,Li:2021zcf}, allowing one to exploit the exceptional angular resolution of tracking detectors at the LHC, and push towards higher-point and higher-precision correlators. 
For recent progress towards understanding the structure of higher-point correlation functions of energy correlators in perturbation theory, see \cite{Chen:2019bpb,Chen:2022jhb,Chang:2022ryc,Yan:2022cye,EECgaussian}.

%%%%%%%%%%%%%%%%%%%%%%%%%%%%%%%%%%%%%%%%%%%%%%%%
%%%%%%%%%%%%%%%%%%%%%%%%%%%%%%%%%%%%%%%%%%%%%%%%
\emph{The Two-Point Correlator $\langle \cE(\vec n_1) \cE( \vec n_2) \rangle$.}---The simplest possible jet substructure observable is the two-point correlator of energy flow operators, $\langle \cE(\vec n_1) \cE( \vec n_2) \rangle $, also referred to as the energy-energy correlator (EEC) \cite{Basham:1979gh,Basham:1978zq,Basham:1978bw,Basham:1977iq}. 
As the most basic correlator, it has been studied extensively in perturbation theory \cite{Belitsky:2013ofa,Dixon:2018qgp,Luo:2019nig,Henn:2019gkr}, and the OPE has been understood non-perturbatively in CFTs \cite{Kologlu:2019mfz,Chang:2020qpj}. 
In addition to controlling the structure of asymptotic energy flow in this theory, this OPE also controls modular flow \cite{Casini:2017roe}. 
While the two-point correlator was measured early on in $e^+e^-$ colliders \cite{SLD:1994idb,L3:1992btq,OPAL:1991uui,TOPAZ:1989yod,TASSO:1987mcs,JADE:1984taa,Fernandez:1984db,Wood:1987uf,CELLO:1982rca,PLUTO:1985yzc} the high energies of the LHC allowed for the first studies of the scaling behavior in the small angle limit \cite{Komiske:2022enw}.\footnote{See \cite{Moult:2018jzp,Moult:2019vou,Gao:2019ojf,Ebert:2020sfi,Li:2021txc} for studies in the Sudakov regime.}

In \Fig{fig:EEC} we plot our analytic calculation of the two-point energy correlator at both leading logarithm (LL) and next-to-leading logarithm (NLL).\footnote{Here we use the logarithm counting appropriate for single logarithmic collinear observables, as opposed to that for double logarithmic observables.} 
We use the NLO hard functions of \cite{Aversa:1988mm,Aversa:1988fv,Aversa:1988vb,Aversa:1989xw,Aversa:1990uv,Jager:2004jh}, the NLO fragmenting jet functions from \cite{Kang:2016mcy,Kang:2016ehg} and the NLO energy correlator jet constants from \cite{Dixon:2019uzg}. 
To focus on the convergence of the perturbative series, as illustrated by the decreasing scale variations from LL to NLL, we have chosen to normalize the analytic results to the data within the perturbative regime. 
Comparisons without this normalization will be given in the next section. 
We compare our results with CMS Open Data extracted from \cite{Komiske:2022enw},\footnote{This data was made publicly available \cite{CERNOpenDataPortal} by the CMS collaboration~\cite{Chatrchyan:2008aa,CMS:OpenAccessPolicy}. The CMS 2011A Open Data~\cite{CMS:JetPrimary2011A},  was simplified into the ``MIT Open Data'' (MOD) format in \Refs{Komiske:2019jim,komiske_patrick_2019_3340205}. It was analyzed using the Energy-Energy Correlators package \cite{EEC_github} in \cite{Komiske:2022enw}. We thank Patrick Komiske for his development and maintenance of this package.} observing excellent agreement. 
The beautiful scaling behavior governed by the light-ray OPE, which is reminiscent of critical phenomena, is clearly visible!

As emphasized above, contrary to the standard lore of the field, no grooming algorithms were needed even in the complex LHC environment. 
This is a remarkable illustration of the universality of the OPE limit in QFT, and highlights the role of formal theory in understanding the correct field theoretic observables for quantifying correlations in energy flux at colliders. 

In the comparison between theory and data in \Fig{fig:EEC}, no unfolding has been done on the CMS Open Data. 
The theoretical advances described in this \emph{Letter} motivate precise experimental measurements of the energy correlator observables. 
The non-standard nature of these observables suggests that new unfolding procedures may be required, such as the recently introduced machine learning (ML) based approaches \cite{Andreassen:2019cjw}, which would be a beautiful union of new ideas from data science and formal theory to better understand QCD at colliders.

%%%%%%%%%%%%%%%%%%%%%%%%%%%%%%%%%%%%%%%%%%%%%%%%
%%%%%%%%%%%%%%%%%%%%%%%%%%%%%%%%%%%%%%%%%%%%%%%%
\begin{figure}
\includegraphics[width=0.955\linewidth]{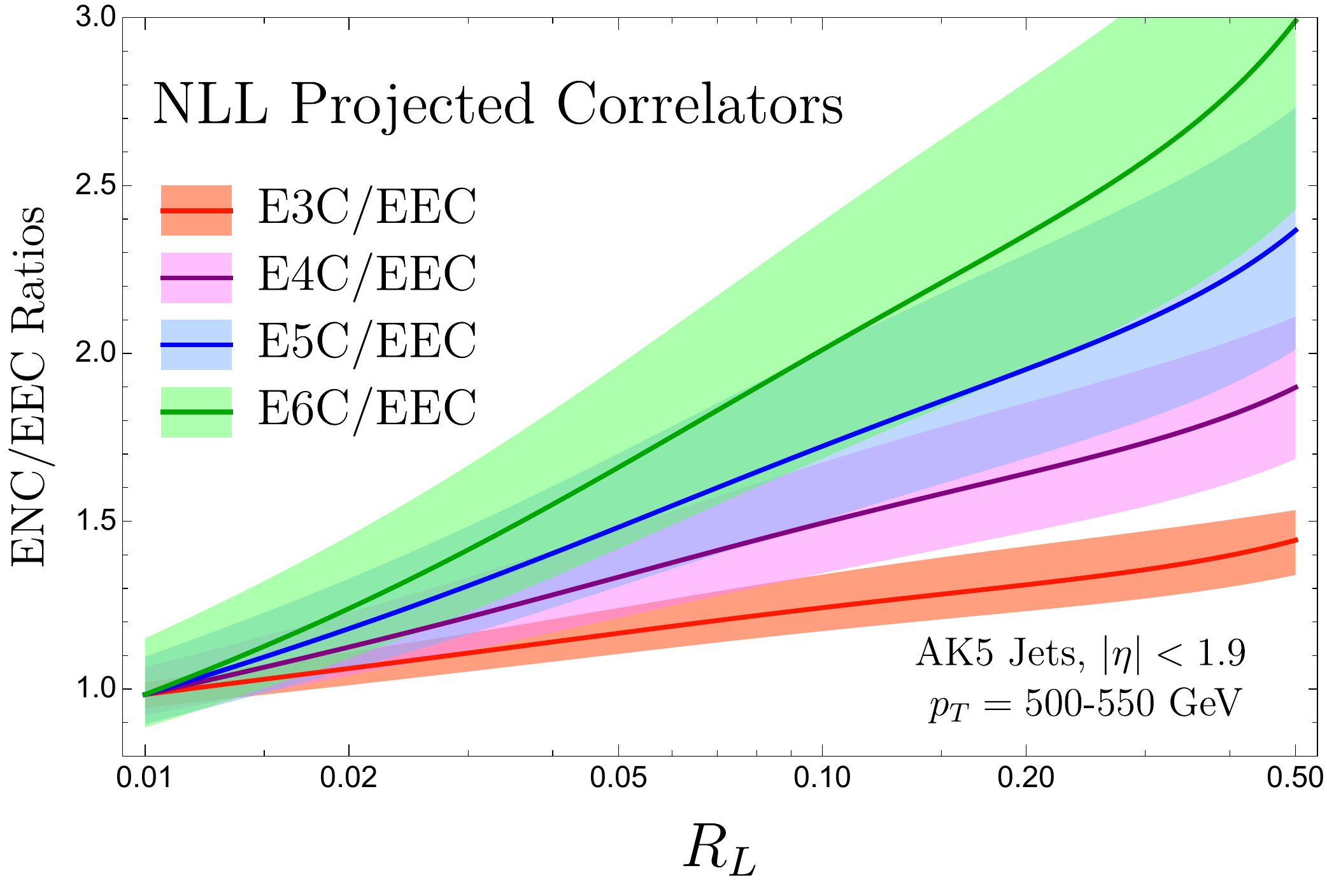}
\caption{Ratios of the multi-point projected correlators at NLL. To focus on the slope, we have identified the central curves at $R_L = 0.01$. 
The theoretical accuracy achieved in this \emph{Letter} allows us to resolve the structure of the quantum scaling dimensions of QCD light-ray operators within high-$p_T$ jets at the LHC!
 }
\label{fig:slope}
\end{figure}

\emph{Higher-Point Projected Correlators.}---In addition to the two-point correlator, we can probe the scaling behavior of higher point correlation functions. 
We consider the projected correlators, ENC$(R_L)$, defined by integrating out the non-trivial shape dependence of the $N$ point correlator, keeping only the longest side $R_L$ fixed \cite{Chen:2020vvp}. 
The projected correlators were first studied in LHC data in \cite{Komiske:2022enw}. 
To isolate the quantum scaling, and suppress non-perturbative corrections, we consider the ratio to the two-point correlator, ENC$(R_L)/$E2C$(R_L)$ \cite{Chen:2020vvp}.

In \Fig{fig:EEC_ratios} we show a comparison of our NLL analytic calculations of the projected correlators, again compared with CMS Open data from \cite{Komiske:2022enw}. 
We find excellent agreement in the perturbative regime. The deviation as $R_L\to 0.5$ is due to boundary effects from the jet clustering. 
In the case that the anomalous dimensions vanished, these curves would be flat.
These ratios probe in detail the anomalous dimensions of the higher spin light-ray operators, and the iterated light-ray OPE! 
The increasing slope occurs due to the monotonic increase of the twist-2 anomalous dimensions as a function of $N$ \cite{Nachtmann:1973mr,Komargodski:2012ek,Costa:2017twz}.
This is well reproduced in the NLL calculation showing sensitivity to the scaling at the quantum level.

The NLL accuracy achieved in this \emph{Letter} is essential for clearly resolving the different scaling behaviors of the $N$-point correlators.
In \Fig{fig:slope} we show all of our NLL predictions in a single plot, and to focus on the scaling, we have identified them at $R_L=0.01$. 
With the reduced uncertainty bands at NLL, we are able to begin to clearly separate the scaling behaviors. 
This is not possible at LL due to the much larger uncertainty bands (see e.g. \Fig{fig:EEC} as representative).
This strongly motivates extending our calculation to NNLL to further reduce the theoretical uncertainties, as well as performing a proper experimental analysis with robust error bars.
Both of these goals are achievable in the near future, and will allow us to resolve the structure of the quantum scaling dimensions of QCD within high-$p_T$ jets!

Our results are particularly interesting for comparing with implementations of higher order DGLAP in parton showers \cite{Hoche:2017iem,Hoche:2017hno,Dulat:2018vuy,Gellersen:2021eci,Li:2016yez}, since the projected correlators cleanly probe the low spin DGLAP anomalous dimensions in a way that has not been possible previously.
Such ``reference resummations" will play an essential role in the development of improved parton showers (see e.g. \cite{Li:2016yez,Hoche:2017hno,Hoche:2017iem,Dulat:2018vuy,Gellersen:2021eci,Hamilton:2020rcu,Dasgupta:2020fwr,Hamilton:2021dyz,Karlberg:2021kwr} for recent work), which will in turn extend our ability to search for new phenomenon at the LHC.

%%%%%%%%%%%%%%%%%%%%%%%%%%%%%%%%%%%%%%%%%%%%%%%%
%%%%%%%%%%%%%%%%%%%%%%%%%%%%%%%%%%%%%%%%%%%%%%%%
\emph{Conclusions.}---In this \emph{Letter} we have bridged the gap between a wealth of developments in the study of conformal collider physics and real world LHC phenomenology. 
This was achieved by deriving a rigorous factorization theorem for the light-ray density matrix at the LHC, in which the energy correlators observables can be evaluated. 
We computed this density matrix at NLO, and combined it with the NLL scaling of the light-ray operators, matching the state of the art accuracy for jet substructure at the LHC, but for a family of observables. 
To our knowledge, this is the first factorization theorem for an infrared safe jet substructure observable at the LHC, that is based on a rigorous QCD factorization theorem in QCD, and absent non-global logarithms, including in the normalization. 

Although we have focused in this \emph{Letter} on applications to the LHC, the interface between energy correlators and fragmenting jet functions, described by the factorization theorem in \Eq{eq:fact_theorem}, can also be applied in heavy ion colliders, or at the future electron-ion collider. 
This will be of interest for precision studies of hadronization, cold nuclear matter, and the quark-gluon plasma using jet substructure.

We believe that our results are of interest both phenomenologically and theoretically. 
On the phenomenological side, a key advantage of our formulation of jet substructure is its simplicity and lack of grooming algorithms, clearing the path towards higher order calculations and the ultimate hope of \emph{precision} QCD phenomenology at the LHC. 
On the theoretical side, our results enable the experimental study of correlation functions of light-ray operators beyond the leading order.  
We hope that this combination can drive genuine progress in our understanding of jet substructure, leading to new insights into QCD, and ever more sophisticated ways to search for new physics and new interactions at the LHC.

%%%%%%%%%%%%%%%%%%%%%%%%%%%%%%%%%%%%%%%%%%%%%%%%
%%%%%%%%%%%%%%%%%%%%%%%%%%%%%%%%%%%%%%%%%%%%%%%%
\emph{Acknowledgements.}---We thank Cyuan-Han Chang, Hao Chen, Patrick Komiske, Jingjing Pan, David Simmons-Duffin, Jesse Thaler, Feng Yuan, Hua Xing Zhu for helpful comments, discussions, questions, and collaboration on related work. 
We thank Felix Ringer for useful discussions and collaboration at the initial stages of the project. 
K.L. is supported by the LDRD program of LBNL, the U.S. DOE under contract number DE-AC02-05CH11231. 
B.M. and I.M. are supported by start-up funds from Yale University. 
I.M. thanks the KITP Santa Barbara for hospitality while parts of this work were completed.
%%%%%%%%%%%%%%%%%%%%%%%%%%%%%%%%%%%%%%%%%%%%%%%%
%%%%%%%%%%%%%%%%%%%%%%%%%%%%%%%%%%%%%%%%%%%%%%%%
\bibliography{EEC_LHC.bib}{}
\bibliographystyle{apsrev4-1}

\end{document}